\documentclass[pre,preprint,english,showpacs,aps,prb,a4paper,floatfix]{revtex4}
\usepackage[T1]{fontenc}
\usepackage{color}
\usepackage[latin1]{inputenc}
\usepackage{graphicx}
\usepackage{bm}
\usepackage{epsfig}

\begin{document}
\title{Liquid-crystal patterns of rectangular particles in a square nanocavity}
\author{Miguel Gonz\'alez-Pinto}
\email{miguel.gonzalezp@uam.es}
\address{ Departamento de F\'{\i}sica Te\'orica de la Materia Condensada, Facultad de Ciencias, Universidad Aut\'onoma de Madrid,
E-28049 Madrid, Spain}
\author{Yuri Mart\'{\i}nez-Rat\'on}
\email{yuri@math.uc3m.es}
\address{Grupo Interdisciplinar de Sistemas Complejos (GISC), Departamento de Matem\'aticas,
Escuela Polit\'ecnica Superior, Universidad Carlos III de Madrid,
Avenida de la Universidad 30, 28911 Legan\'es, Madrid, Spain}
\author{Enrique Velasco}
\email{enrique.velasco@uam.es}
\address{Departamento de F\'{\i}sica Te\'orica de la Materia Condensada and  Instituto de F\'{\i}sica de la Materia Condensada, 
Facultad de Ciencias, Universidad Aut\'onoma de Madrid,
E-28049 Madrid, Spain}

\begin{abstract}
Using density-functional theory in the restricted-orientation approximation,
we analyse the liquid-crystal patterns and phase behaviour of a fluid of hard
rectangular particles confined in a two-dimensional square nanocavity of side length $H$ composed of hard inner walls. 
Patterning in the cavity is governed by surface-induced order, capillary and frustration effects, and 
depends on the relative values of particle aspect ratio $\kappa\equiv L/\sigma$, with $L$ the length and $\sigma$ the width of the 
rectangles ($L\ge\sigma$), and cavity size $H$. Ordering may be very different from bulk ($H\to\infty$) behaviour when $H$ is
a few times the particle length $L$ (nanocavity). Bulk and confinement properties are obtained for the cases $\kappa=1$, $3$ and 
$6$. In bulk the isotropic phase is always stable at low packing fractions $\eta=L\sigma\rho_0$ (with
$\rho_0$ the average density), and nematic, smectic, columnar and crystal phases can be stabilised at higher $\eta$ depending on
$\kappa$: for increasing $\eta$ the sequence isotropic $\to$ columnar is obtained for $\kappa=1$ and $3$, whereas 
for $\kappa=6$ we obtain isotropic $\to$ nematic $\to$ smectic (the crystal being unstable in all three cases for the 
density range explored). In the confined fluid surface-induced frustration leads to
four-fold symmetry breaking in all phases (which become two-fold symmetric). 
Since no director distorsion can arise in our model by construction, frustration in the director orientation
is relaxed by the creation of domain walls (where the director changes by $90^{\circ}$); this configuration is
necessary to stabilise periodic phases. For $\kappa=1$ the crystal becomes 
stable with commensuration transitions taking place as $H$ is varied. These transitions involve structures with
different number of peaks in the local density. In the case $\kappa=3$ the commensuration transitions involve
columnar phases with different number of columns. Finally, in the case $\kappa=6$,
the high-density region of the phase diagram is dominated by commensuration transitions between smectic structures;
at lower densities there is a symmetry-breaking isotropic $\to$ nematic transition exhibiting non-monotonic behaviour
with cavity size. Apart from the present application in a confinement setup, our model could be used to explore the bulk
region near close packing in order to elucidate the possible existence of disordered phases at close packing.
\end{abstract}

\maketitle

\section{Introduction}

Confinement is known to affect the liquid-crystal ordering behaviour of a fluid in a dramatic fashion.
In fluids with first-order isotropic (I) to nematic (N), I$\to$N, transitions in bulk,
there exists a corresponding I$\to$N transition when the fluid is confined into a pore of size $h$. 
Depending on the nature of the liquid crystal, the transition is shifted 
by $\Delta T$ in temperature $T$, or by $\Delta\mu$ in chemical potential $\mu$, with respect to the bulk transition, 
a phenomenon called {\it capillary ordering}. This effect is similar to that occurring in normal, isotropic liquids 
(the so-called {\it capillary condensation}), and is governed by the corresponding macroscopic Kelvin
equation \cite{Sluckin}, which predicts $\Delta T$ or $\Delta\mu\sim h^{-1}$, for $h\to\infty$. The nature of surface interactions determine
whether the shift is positive or negative and $O\left(h^{-2}\right)$ and higher-order corrections may be
included for not-so-large pores. This phenomenon also applies to liquid-crystalline phases
exhibiting partial or full spatial order, such as smectic (S), columnar (C) or crystal (K), provided the bulk 
transition is of first-order, and the behaviour of the confined transition is described by a corresponding Kelvin 
equation, valid in the r\'egime of large pores, with corrections due to elasticity of the periodic structure.

Confinement brings about further complex phase 
behaviour due to commensuration effects between the cavity size $h$ and the natural periodicity of the liquid-crystal 
structure, $a$, in the case of partially- or fully-ordered phases. In some cases the ordered structure can be suppressed 
altogether when $h\ne na$, where $n$ is an integer. This effect has been studied in crystals
\cite{Guillermo} and in liquid crystals and similar systems in three \cite{us,us1,polish,polish1,binder} and two \cite{Yuri} 
dimensions; in all of these cases,
{\it commensuration transitions} are observed involving structures with different numbers of unit cells. In the case of 
S phases these transitions are sometimes called layering transitions and are similar to those occurring in adsorption
systems since they involve the creation of a new layer in a step-wise manner. In some circumstances the 
capillary-ordering transition interacts with the commensuration transitions in a complicated manner and, for small pore 
sizes, capillary ordering may become intermittent.

In this paper we focus on the additional phenomenology caused by the presence of surface-induced frustration in a 
fluid confined in a nanocavity, i.e. a cavity a few particle lengths in size. Surfaces are known to be able to orient 
(anchor) the director of a liquid crystal along specific directions, meaning that
the fluid will pay a surface free-energy cost for not choosing the favoured alignment at the wall so that any
gross misalignment is discouraged (strong anchoring condition). If two regions of 
the confining surface favour different directions, the liquid crystal will be subject to frustration: distortion (or
elastic) and defect free energies, both of which cannot be optimised at the same time, will compete. In turn, 
due to capillarity and commensuration effects, the resulting phase behaviour, involving typically the formation of 
defects, may indeed be quite complex. 

\begin{figure}
\includegraphics[width=3.0in,angle=0]{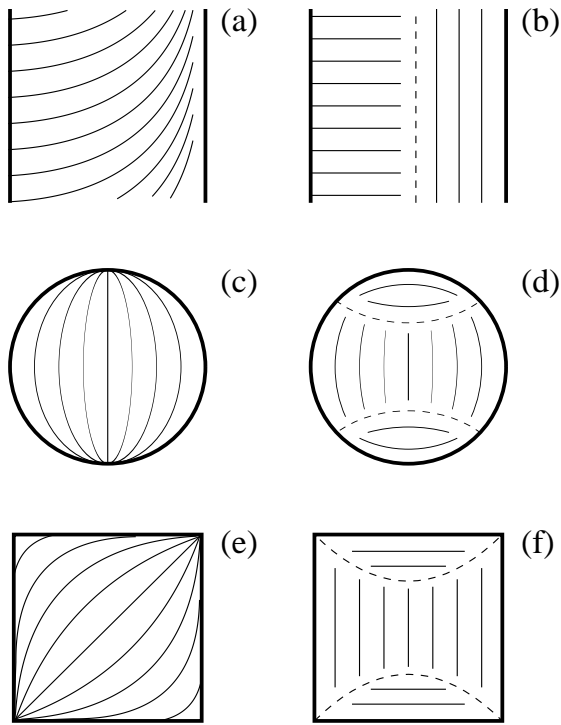}
\caption{Schematic of three 2D systems with surface-induced frustration. 
(a) and (b): planar slit pore with favoured surface orientation of the director at right angles.
(c) and (d): circular cavity with planar surface orientation. (e) and (f): square cavity with planar surface
orientation. In all cases surface free energy is satisfied, but in (a), (c) and (e) the director field is distorted
with possible creation of point defects, whereas in (b), (d) and (f) domain walls separating nematic
regions with different orientations (dashed lines) are formed.}
\label{fig1}
\end{figure}

A simple geometry for the surface-induced frustration effect is a planar slit-pore system with the two
walls favouring different alignments, for example, homeotropic ($\psi=0^{\circ}$, where $\psi$ is the angle between 
the nematic director
and the surface normal) and planar ($\psi=90^{\circ}$). This system is depicted schematically in Fig. \ref{fig1}(a) and
(b). In the nematic r\'egime, one of two situations may happen: either the nematic director smoothly rotates between the 
two surfaces, incurring an elastic free energy [distorted configuration, Fig. \ref{fig1}(a)], 
or a planar interface separating two films of uniform nematic with two competing 
orientations may be formed [Fig. \ref{fig1}(b)], called a step-like phase. A structural transition between the two structures arises as pore size 
or thermodynamic conditions are varied. This phenomenon was predicted to occur in the region about a line defect \cite{Sluckin1}. In a slit pore, it was
predicted using Landau-de Gennes theory \cite{Palffy,LdG}, and has been further analysed
with the same technique \cite{slovenos}, by density functional theory \cite{us_step,Teix} and by simulation \cite{italian1,italian2,us_LL}.
The step-like phase is due to strong anchoring conditions of the director at the surfaces together with a large elastic free energy
compared to the free energy of formation of step interfaces or walls.  
A similar phenomenon has been studied, using Monte Carlo simulation, by de las Heras and Velasco \cite{HerasVelasco} in 
two-dimensional fluids of hard rods
confined in circular cavities inducing planar anchoring conditions, where the circular wall induces a distorted 
director configuration with two point defects at intermediate densities, Fig. \ref{fig1}(c); at higher densities two 
interfaces separating nematic regions with a slight director distortion is a more stable configuration, 
Fig. \ref{fig1}(d). 
In a square cavity the distorted configuration will have four defects at the corners, Fig. \ref{fig1}(e), which will
give rise at higher densities to an undistorted phase with two domain walls, Fig. \ref{fig1}(f). A crucial point is that 
spatially nonuniform phases such as S, C or K phases, may only form, in nanocavities, in undistorted director configurations,
so that the stabilisation of the domain-wall configurations depicted in Figs. \ref{fig1}(b), (d) or (f) is a necessary 
condition for the formation of these nonuniform phases. Once the domain-wall phases have been stabilised the
formation of nonuniform phases and the occurrence of commensuration transitions is possible.   

{The way some regions of the confined C, S or K phases with different director orientation and/or number of 
layers grow at the expense of the others when external conditions are changed is an interesting problem. The topology of the free-energy 
landscape defined by all the local minima (whether stable or not) separated by energy barriers
is crucial as it dictates the path the system will follow between two equilibrium states. A study of the dynamical evolution
would require to impose mass conservation to derive the evolution equations for the density profiles. Based on the dynamical 
and equilibirum density functional theory, a phase-field liquid-crystal model was recently derived 
which allows the study of nonequilibrium density and order parameter dynamic evolutions \cite{Lowen}.  
However, the evolution that follows from the conjugate-gradient scheme used in the present work to obtain the possible local minima of the grand 
potential as the state variables are changed will result in a `time' evolution that resembles the real dynamical evolution.
We have used this technique to analyse how the textures of the confined C, S or K phases (depending on the value of particle aspect ratio)
evolve between two states with different number of cells (columns, layers or nodes) as the external conditions is varied. The mechanism by which
new periodic cells grow depends on the symmetry of the confined phases (particles oriented parallel, C, or perpendicular, S, to the layers),
on the nature (continuous versus first order) of the bulk phase transitions, and on the amount of orientational ordering of the system.}  

The interest in these studies is also motivated by the possibility that experiments on quasiamonolayers of vibrated 
granular rods can be related to thermally equilibrated molecular or colloidal fluids of hard anisotropic particles 
\cite{Aranson,indios,Galanis}. There is ample evidence that
these monolayers do form liquid-crystalline phases, and surface effects, nematic ordering, etc. may be explored in these
experimental systems and compared with their thermal counterparts to look for similarities in their ordering behaviour, especially   
considering that overlap or exclusion interactions are the key interactions in both systems.

In this paper we use a simple fundamental-measure density-functional theory to study the ordering properties and thermodynamic
behaviour of a fluid of hard rectangular particles confined in a square cavity composed of hard inner walls.
The model only admits two particle orientations and is unable to describe configurations where the director field
is distorted. Since we are interested in the interplay between surface-induced frustration, capillarity and
commensuration as far as high-density, nonuniform phases are concerned, the nonexistence of the intermediate 
configuration depicted in Fig. \ref{fig1}(e) is not problematic. The model correctly describes spatial correlations 
between rectangles in parallel or perpendicular configurations and therefore is expected to give correct 
predictions on the liquid-crystal patterns and thermodynamic phase behaviour of the system at high densities.   
Irrespective of the length-to-width ratio of the particles, all liquid-crystal structures found have the two-fold
symmetric configuration depicted in Fig. \ref{fig1}(f) where the different regions are fluid (N phase) or organised 
into layers (S phase) or columns (C phase), depending on the elongation of the rectangles; for squares the four-fold
symmetry of the cavity is conserved and the high-density phase is a K phase on a square lattice. The bulk and
confinement phase diagrams were calculated for a number of systems. The high-density r\'egime of the latter are
dominated by commensuration transitions between periodic (S, C or K depending on the system) structures involving
different number of periodic cells. Except for details on the phase diagram and the mechanisms involved in these
commensuration transitions, the phase behaviour of this fluid under confinement seems to be quite universal.

The paper is arranged as follows. In the following section the particle model and system are introduced. Then, in
Section \ref{FE}, the density-functional theory used is defined, and we discuss how the bulk and confined phases
are obtained numerically. The results are presented in Section \ref{Results}, which is divided in subsections,
one for each fluid considered. Section \ref{limitations} is devoted to discussing the limitations of the model,
together with some ways to improve the model and extend it to study the pattern growth dynamics associated with
the commensuration transitions. Finally in Section \ref{conclusions} we summarise the work and present some conclusions.

\section{Particle model and system}

\begin{figure}[h]
\includegraphics[width=2.5in,angle=0]{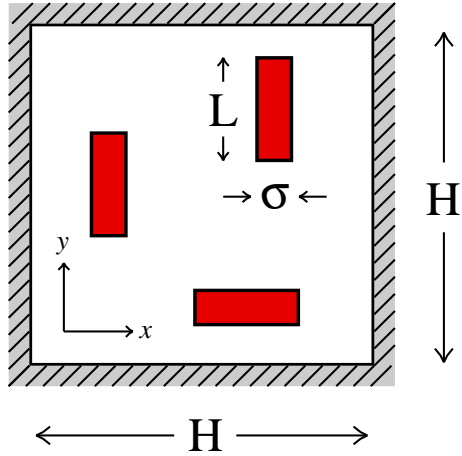}
\caption{Schematic of the system studied: a fluid of hard rectangles of length $L$ and width $\sigma$ that can orient only
along two perpendicular directions $x$ and $y$, inside a square cavity of side length $H$.}
\label{system}
\end{figure}

The particle model used is a rectangle of sizes $L$ (length) and $\sigma$ (width), Fig. \ref{system}. The $\sigma$ 
parameter will be
used as length scale and we adopt the packing fraction $\eta=L\sigma\rho_0$ as the scaled density parameter; $\rho_0=N/A$ 
is the mean density ($N$ is the number of particles and $A$ the area). The size of the particles will
be defined in terms of their {\it aspect ratio}, $\kappa\equiv L/\sigma$. Particles interact with overlap or exclusion 
interactions [hard rectangle (HR) model], so that the temperature $T$ is irrevelant as far as the phase behaviour of the 
system is concerned. In our model we will use the restricted-orientation or Zwanzig approximation, where particles can 
only point along four directions, say the $x$ and $y$ axes in the positive and negative directions. 
At low $\eta$ a fluid of hard rectangles is in the I phase,
with the long axes of particles pointing in all directions with equal probability. As the packing fraction
$\eta$ is increased, the fluid becomes oriented (N phase) if $\kappa\agt 3$ (the precise value 
has not been determined yet). In the high-density r\'egime phases with spatial order, S, C or K
can be stabilised. Low values of $\kappa$ favour the C phase where particles arrange into columns (consisting of
a one-dimensional `fluid' of particles flowing in the direction of the long particle axes), whereas for higher $\kappa$ 
the S phase (with `fluid' rows of particles stacked one on top of the other) tends to be more stable (note that for the special 
case $\kappa=1$ the C and S phases are identical). The stable phase at densities near the close-packing value 
$\eta=1$ [or $\rho_0=(L\sigma)^{-1}$] is not known ({except for hard squares where a free-energy 
minimization using a Gaussian parameterization 
for the local density predicts the K to be the stable phase}), but entropic effects might favour 
a partially fluid (S or C) or disordered K phase against the expected, fully-ordered K phase.

Several theoretical \cite{Schlacken,MRaton} and computer-simulation \cite{Wojciechowski,Donev} studies of the HR fluid,
both in the restricted- and free-orientation cases, exist. These studies were mainly motivated by the prediction that
this fluid can exhibit an exotic nematic, the so-called tetratic phase, possessing four-fold rotational symmetry
(even though the particles only have two-fold symmetry). The tetratic phase, which has been observed in colloidal
fluids \cite{Chaikin}, is present at low values of $\kappa$,
but there are indications that tetratic correlations persist up to substantial values of $\kappa$ 
\cite{YuriKike,Mederos,Selinger}. For large $\kappa$ the usual I$\to$N transition is expected at low density. The high-density
fluid has only been studied so far for aspect ratios $\kappa=1$ \cite{Wojciechowski,Hoover}, $2$ \cite{Donev} and $3$ \cite{Yuri}.
As mentioned before, the
existence of a perfect crystal is questionable because rectangles will perfectly pack even in disordered arrangements,
so that a residual entropy might exist at close packing. Experiments on hard square colloids \cite{Zhao} observe a 
striking transition from a hexagonal rotator crystal to a rhombic crystal via a first-order transition. Simulations on 
parallel hard squares \cite{Hoover} obtain a continuous transition to a K phase at $\eta=0.79$. 
{A free-minimization of the fundamental-measure density-functional theory for hard squares \cite{vanRoij}  
(identical to that used in the present work) gives a C$\to$K first-order transition at $\eta\sim 0.73$; both phases 
bifurcate from the I branch at $\eta=0.538$.}
In the case $\kappa=2$ simulations
have shown the possibility that the high-density phase consists of a nonperiodic, random tetratic, possibly glassy
phase with a residual entropy of $1.79k$ per particle (where $k$ is Boltzmann's constant). For $\kappa=3$ 
the observed high-density phase in the
density-functional study of parallel HR by Mart\'{\i}nez-Rat\'on \cite{Yuri} is the C phase.
C-like layering was observed in simulations of a HR fluid confined in a two-dimensional slit pore \cite{Triplett}, 
and layering transitions between C phases with different number of layers were obtained in Ref. \cite{Yuri} using
density-functional theory.

\section{Free-energy functional}
\label{FE}

The free-energy density functional used is a fundamental-measure functional for hard rectangles 
where only two particle orientations, along $x$ and $y$, are possible. This can be mapped onto the equivalent
problem of a mixture of two components of local densities $\rho_{\nu}({\bm r})$, with $\nu=\{x,y\}$ and where 
${\bm r}$ refers to the position of the particle centres of mass. As usual, the Helmholtz free-energy functional 
${\cal F}[\rho_{\nu}]={\cal F}_{\hbox{\tiny id}}[\rho_{\nu}]+{\cal F}_{\hbox{\tiny ex}}[\rho_{\nu}]$
is split into ideal ${\cal F}_{\hbox{\tiny id}}[\rho_{\nu}]$ and excess 
${\cal F}_{\hbox{\tiny ex}}[\rho_{\nu}]$ parts, with
\begin{eqnarray}
\beta{\cal F}_{\hbox{\tiny id}}[\rho_{\nu}]=\sum_{\nu}\int_Ad{\bm r}\rho_{\nu}({\bm r})
\left\{\log{\left[\rho_{\nu}({\bm r})\Lambda_{\nu}^3\right]}-1\right\},
\end{eqnarray}
where $\beta=1/kT$, $T$ is the temperature, $A$ the system area, and $\Lambda_{\nu}$ the thermal 
wavelength of the $\nu$-th species. In fundamental-measure theory, the excess free-energy functional is 
written as $\beta{\cal F}_{\hbox{\tiny ex}}[\rho_{\nu}]=\int_Ad{\bm r}\Phi({\bm r})$, where $\Phi({\bm r})$ is 
the following local free-energy density {\cite{Cuesta}}:
\begin{eqnarray}
\Phi({\bm r})=-n_0({\bm r})\log{\left[1-n_2({\bm r})\right]}+\frac{n_{1x}({\bm r})
n_{1y}({\bm r})}{1-n_2({\bm r})}.
\end{eqnarray}
Here $n_{\alpha}({\bm r})$ are weighted densities defined as convolutions:
\begin{eqnarray}
n_{\alpha}({\bm r})=\sum_{\nu}\int_A d{\bm r}'\rho_{\nu}({\bm r}')\omega_{\nu}^{(\alpha)}({\bm r}-{\bm r}').
\end{eqnarray}
The weighting functions $\omega_{\nu}^{(\alpha)}({\bm r})$ are particle geometrical measures related to
{the corners, edge-lengths and surface of a rectangle}:
\begin{eqnarray}
&\displaystyle\omega_{\nu}^{(0)}({\bm r})=\frac{1}{4}\delta\left(\frac{\sigma_x^{\nu}}{2}-\left|x\right|\right)
\delta\left(\frac{\sigma_y^{\nu}}{2}-\left|y\right|\right),&\displaystyle
\omega_{\nu}^{(1x)}({\bm r})=\frac{1}{2}\Theta\left(\frac{\sigma_x^{\nu}}{2}-\left|x\right|\right)
\delta\left(\frac{\sigma_y^{\nu}}{2}-\left|y\right|\right),\nonumber\\\nonumber\\
&\displaystyle\omega_{\nu}^{(1y)}({\bm r})=\frac{1}{2}\delta\left(\frac{\sigma_x^{\nu}}{2}-\left|x\right|\right)
\Theta\left(\frac{\sigma_y^{\nu}}{2}-\left|y\right|\right),&\displaystyle
\omega_{\nu}^{(2)}({\bm r})=\Theta\left(\frac{\sigma_x^{\nu}}{2}-\left|x\right|\right)
\Theta\left(\frac{\sigma_y^{\nu}}{2}-\left|y\right|\right),
\end{eqnarray}
with $\delta(x)$ and $\Theta(x)$ the Dirac-delta and Heaviside functions, respectively. We have defined
$\sigma_{\mu}^{\nu}=\sigma+(L-\sigma)\delta_{\mu\nu}$, with $\delta_{\mu\nu}$ the Kronecker symbol
(note that $\sigma_{\mu}^{\mu}=L$ and $\sigma_{\mu}^{\nu}=\sigma$ if $\mu\ne\nu$).
  
The fluid is confined into a hard square cavity of side length $H$ which can be described in terms of an 
external potential $V_{\hbox{\tiny ext}}^{(\nu)}({\bm r})$. The potential acts as an impenetrable wall on 
the particles, i.e.
\begin{eqnarray}
\beta V_{\hbox{\tiny ext}}^{(\nu)}({\bm r})=\left\{\begin{array}{cc}\displaystyle
0,&\displaystyle\frac{\sigma_x^{\nu}}{2}<x< H-\frac{\sigma_x^{\nu}}{2}\hbox{ and }
\frac{\sigma_y^{\nu}}{2}<y<H-\frac{\sigma_y^{\nu}}{2},\\\\
\infty,&\hbox{otherwise.}\end{array}\right.
\end{eqnarray}
The grand potential of the fluid is then:
\begin{eqnarray}
\Omega[\rho_{\nu}]={\cal F}[\rho_{\nu}]-\sum_{\nu}\int_Ad{\bm r}\rho_{\nu}({\bm r})\left[\mu_{\nu}-
V_{\hbox{\tiny ext}}^{(\nu)}({\bm r})\right],
\end{eqnarray}
where $\mu_{\nu}$ is the chemical potential of the $\nu$-th species.
The equilibrium state of the system is obtained by minimising the grand potential at fixed values of 
$\mu_{\nu}$.

From the equilibrium local densities $\rho_{\nu}({\bm r})$ it is possible to define the total local
density 
\begin{eqnarray}
\rho({\bm r})=\sum_{\nu}\rho_{\nu}({\bm r})=\rho_x({\bm r})+\rho_y({\bm r})
\label{pt}
\end{eqnarray}
and a local packing fraction $\eta({\bm r})=\rho({\bm r})L\sigma$.
A local order parameter $Q({\bm r})$ can also be defined from the local ordering tensor, 
with elements $Q_{ij}=\left<2\hat{\bm e}_i\hat{\bm e}_j-\delta_{ij}\right>$. Here
$\left<\cdots\right>$ is a thermal average and $\hat{\bm e}$ is the particle long
axis, with ${i,j=1,2}$ denoting the two Cartesian components. Since, in our restricted-orientation model, 
the local angular distribution function $h(\varphi,{\bm r})$ can be written as
\begin{eqnarray}
h(\varphi,{\bm r})=\frac{\rho_x({\bm r})}{\rho({\bm r})}\left[\frac{\delta\left(\varphi\right)+\delta\left(\varphi-\pi\right)}{2}\right]+
\frac{\rho_y({\bm r})}{\rho({\bm r})}\left[\frac{\delta\left(\varphi-\frac{\pi}{2}\right)+\delta\left(\varphi-\frac{3\pi}{2}\right)}{2}
\right],
\end{eqnarray}
the elements of the local ordering tensor are
\begin{eqnarray}
Q_{ij}({\bm r})=\int_0^{2\pi}d\varphi h(\varphi,{\bm r})\left(\begin{array}{cc}
\cos{2\varphi}& \sin{2\varphi}\\
\sin{2\varphi}& -\cos{2\varphi}\end{array}\right)=Q({\bm r})\left(\begin{array}{cc}
1&0\\0& -1\end{array}\right).
\end{eqnarray}
Then the local order parameter is
\begin{eqnarray}
Q({\bm r})=\int_0^{2\pi}d\varphi \cos{2\varphi}h(\varphi,{\bm r})=
\frac{\rho_x({\bm r})-\rho_y({\bm r})}{\rho({\bm r})}.
\label{Qr}
\end{eqnarray}
Clearly $-1\le Q\le +1$. Since the ordering tensor is always diagonal in the reference frame 
defined by the two possible perpendicular orientations, in the present model the director 
can only point along two directions: the $x$ axis ($Q>0$) or the $y$ axis ($Q<0$). 
Intermediate situations with the director pointing at an angle $\varphi$ different from
$0^{\circ}$, $90^{\circ}$, $180^{\circ}$ or $270^{\circ}$
cannot be described. In particular, a nematic film sandwiched between two uniform parallel lines that exert 
antagonistic boundary conditions [e.g. parallel and homeotropic, Fig. \ref{fig1}(b)] will always develop a step or 
domain wall since the director cannot rotate to form a linearly rotating director field (the stable solution predicted
by elasticity theory). Also, distorted-director configurations such as that represented in Fig. \ref{fig1}(e) for
the square cavity will never appear as solutions of the model. Still within a restricted-orientation approximation,
at least a few intermediate orientations, e.g. $\varphi=45^{\circ}$ and $135^{\circ}$ (and, due to the particle
head-tail symmetry, their equivalent orientations $\varphi=225^{\circ}$ and $315^{\circ}$) would be needed to describe
such configurations. Since we are mainly interested in the high-density r\'egime of the model, we did not pursue that 
line here.

\subsection{Bulk nematic phase and I$\to$N phase transition}

In the case of a uniform nematic phase, $\rho_{\nu}({\bm r})=\rho_{\nu}=$ constant, and the free
energy simplifies considerably. Inverting the equations that define $\rho$ and $Q$ in terms of
$\rho_x$ and $\rho_y$, we obtain
\begin{eqnarray}
\rho_x=\frac{\rho}{2}\left(1+Q\right),\hspace{0.4cm}
\rho_y=\frac{\rho}{2}\left(1-Q\right).
\end{eqnarray}
It is easy to obtain the relations $n_0=\rho$, $n_2=\eta$, and 
\begin{eqnarray}
n_{1x}=\frac{\rho}{2}\left[L+\sigma+\left(L-\sigma\right)Q\right],\hspace{0.4cm}
n_{1y}=\frac{\rho}{2}\left[L+\sigma-\left(L-\sigma\right)Q\right].
\end{eqnarray}
From here, the free energy density is
\begin{eqnarray}
\frac{\beta\cal F}{A}&&=\rho\left\{\log{\rho}-1+\frac{1}{2}\left[\left(1+Q\right)\log{\left(\frac{1+Q}{2}\right)}
+\left(1-Q\right)\log{\left(\frac{1-Q}{2}\right)}\right]\right.\nonumber\\\nonumber\\&&\left.
-\log{\left(1-\eta\right)}+\frac{\eta}{4(1-\eta)}\left[\left(\kappa+\kappa^{-1}+2\right)-
\left(\kappa+\kappa^{-1}-2\right)Q^2\right]\right\}
\end{eqnarray}
where we have defined $\kappa\equiv L/\sigma$ as the aspect ratio of the rectangle 
(here and in the following the thermal wavelengths $\Lambda_{\nu}$ are assumed to be absorbed in the
chemical potentials $\mu_{\nu}$). Direct minimisation
of ${\cal F}$ with respect to $Q$ leads to the following trascendental equation:
\begin{eqnarray}
Q=\frac{\displaystyle 1-e^{-y\left(\kappa+\kappa^{-1}-2\right)Q}}{\displaystyle 1+
e^{-y\left(\kappa+\kappa^{-1}-2\right)Q}},
\end{eqnarray}
where $y\equiv \eta/(1-\eta)$. For packing fractions $\eta\ge\eta^*$ this equation presents a solution 
$Q\ge 0$. The transition can be shown to be continuous. Assuming $Q\ll 1$ and expanding to lowest order,
we find $\eta^*=2/(\kappa+\kappa^{-1})$. Of course the I$\to$N transition is in some cases preempted by
a transition to a nonuniform phase, as we show below.

\subsection{Bulk nonuniform phases}

In order to obtain the stability regions where possible nonuniform (S, C and K) phases can be stable, 
a numerical minimisation of the grand potential $\Omega$ at fixed $\mu_{\nu}$ is required. Assuming a periodic
structure in $x$ and $y$ with periods $d_x$ and $d_y$, respectively, we chose to perform 
a Fourier expansion of the density profile,
\begin{eqnarray}
\rho_{\nu}({\bm r})=\rho_0\gamma_{\nu}\sum_{n,m=0}^{\infty}\alpha_{nm}^{(\nu)}\cos{(nq_xx)}\cos{(mq_yy)},
\label{rho}
\end{eqnarray}
where $\rho_0$ is the average density, $\gamma_{\nu}$ is the occupancy probability per unit cell of species $\nu$ 
(with $\gamma_x+\gamma_y=1$), and $q_x=2\pi/d_x$ and $q_y=2\pi/d_y$ are wavevectors. The coefficients 
$\{\alpha_{nm}^{(\nu)}\}$ are taken as minimisation
variables together with the periods $d_x$ and $d_y$, and one of the molar fractions $\gamma_{\nu}$, 
and a conjugate-gradient technique was used to minimise the grand 
potential. 
In the case of the S and C phases the number of Fourier components can be drastically reduced since we can restrict
the indices to the form $\alpha^{(\nu)}_{n0}$. 
In practice the sums (\ref{rho}) were truncated using the criterion $\left|\alpha_{nm}^{(\nu)}\right|<10^{-p}$.
For $\eta<0.6$ this criterion could be satisfied with $p=8$ and $10\times 10$ Fourier components; for the highest densities
convergence is a bit poorer and $p$ had to be reduced to $5$ (with the same number of Fourier components). In all cases an 
increase in the number of components did not lead to a significant change in the results. 

\subsection{Confined structures}

In the case of the confined system, the grand potential was discretised in real space, and the two densities 
$\rho_x$, $\rho_y$ at each grid point were taken as independent variables. The grid step size was 
$\Delta x=\Delta y=0.05\sigma$ or 20 
points in a particle width $\sigma$ in the cases $\kappa=3$ and $6$; in the case $\kappa=1$ a finer grid with
$\Delta x=\Delta y=\sigma/60$ was used. As an example, that gives a grid of $(3\times 6\times 20)^2=360^2$ points 
in the case $H=3L$ and $\kappa=6$. Again a conjugate-gradient method, with analytical calculation of the gradient,
was used to perform the minimisations. The minimisation variables used were not the local densities, but 
$\sqrt{\rho_{\nu}({\bm r})}$,
since these variables show better convergence properties due to their larger variation close to the regions
where the density is very small. The iterative process is very efficient and the convergence is such that
the absolute value of the conjugate gradient achieved is typically $10^{-12}$ per mesh point. 
Typically the starting configurations were isotropic configurations ($\rho_x=\rho_y$) at a low value of chemical potential. 
After equilibration, $\mu$ was increased in small steps, using the converged densities of the previous step. 
In addition, a series of runs where $\mu$ was decreased at fixed $H$ were performed
to search for discontinuous phase transitions. Runs with varying cavity sizes (increasing and decreasing $H$)
at fixed $\mu$ were also done by changing the cavity size by a few grid points and rescaling the density fields on 
the grid. 

  
\begin{figure}[h]
\includegraphics[width=5in,angle=0]{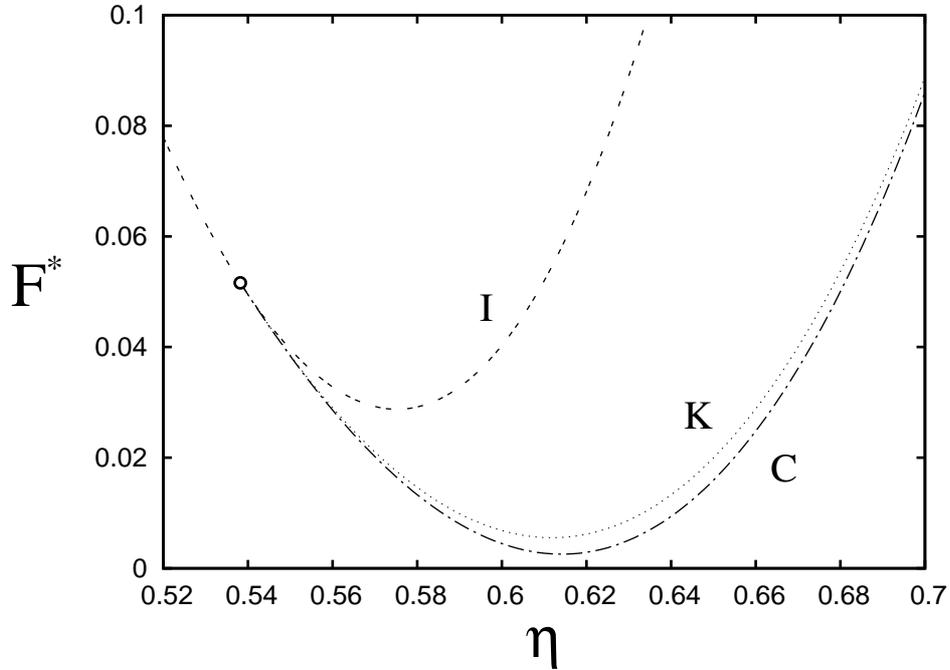}
\caption{{Free energy density branches in reduced units $F^*=\beta {\cal F}L\sigma/A$} 
of the I, C and K phases with respect to packing fraction for the HR fluid with aspect 
ratio $\kappa=1$ (hard squares). The straight line $f=a\eta+b$, with $a=6.20$ and $b=-3.22$, has been 
subtracted from the free energies in order to better visualise the curves. The transition point is located 
at $\eta=0.538$.}
\label{k1}
\end{figure}

\section{Results}
\label{Results}

In this section we show the results of our density-functional investigation on the bulk and confinement phase behaviour 
of the HR fluid. The section is divided into subsections, each including the results for the cases $\kappa=1$, $3$
or $6$.
  
\subsection{$\kappa=1$}

The bulk phase diagram in this case exhibits the sequence 
\begin{eqnarray*}
\hbox{I}\to\hbox{C}\to\hbox{K}.
\end{eqnarray*}
The I$\to$C transition occurs at $\eta=0.538$, and is continuous. 
The K and C phases have very similar free energies, but a K phase with a square-lattice structure becomes more stable at 
a first-order transition occurring at $\eta\sim 0.750$ (this result has been obtained using a Gaussian parameterisation for 
the local densities; 
a parameterisation-free calculation lowers this number by a few hundredths. van Rooij et al. \cite{vanRoij}, 
using the same density functional obtain $\eta\sim 0.73$).  
Fig. \ref{k1} shows the free energies
of the I, C and K phases with respect to packing fraction (note that, in the case $\kappa=1$, the two species $\nu=\{x,y\}$
are degenerate and the system can be viewed as a one-component system rather than as a mixture; in the present results 
the former view was adopted, so that the free energy and chemical potential do not include the factor $-\log{2}$ from the
entropy of mixing). The C and K free-energy branches bifurcate from the I
branch at the same packing fraction (indicated by circles), but the C phase is more stable up to the transition
to the K phase. The transition is continuous within the accuracy of our calculations.
 
\begin{figure}[h]
\includegraphics[width=6in,angle=0]{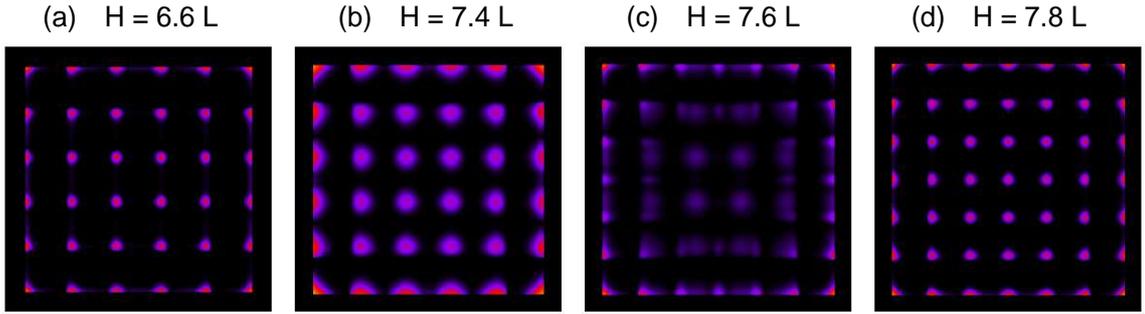}
\caption{Density contour plots of confined structures in the case $\kappa=1$ and for chemical potential
$\beta\mu=8.376$. Different cavity sizes from
$H=6.6L$ to $7.8L$ are shown. These structures correspond to the K$_6\to$K$_7$ transition. Colour saturation has
been adjusted differently in each panel to optimise contrast.}
\label{evol1}
\end{figure}

In the interval of packing fractions explored here, the C phase is marginally more stable than the K phase in bulk.
However, in the confined fluid, the near-degeneracy of the C and K phases in terms of free energy is broken, and the
K phase becomes much more stable; now the restricted geometry and the boundary conditions favour the formation of
well localised density peaks. The reason is the following: {since the cavity has a square shape and particles
are completely symmetric (same length and width), the system cannot break the symmetry along a single direction. Even though 
the four walls could in principle favour C ordering, this situation obviously generates frustration as 
particles cannot freely diffuse within each layer due to their orthogonal intersections. 
Thus the localization of particles on a square lattice minimizes the free energy as 
the total density is increased.}  

The number and configuration of the crystal peaks depend very much on the cavity side
length $H$. If $a$ is the lattice parameter of the K phase (square lattice), we expect $a\simeq 1.18L$ in bulk
(the prefactor does have a slight dependence on chemical potential).
Then a square lattice with $n^2$ peaks ($n$ being an integer) will fit into the cavity when the cavity side is
$H=H_n$, with $H_n=(n-1)a+L$ or $H_n/L\simeq 1.18n-0.18$. Therefore we expect a transition between a structure 
with $n^2$ peaks, labelled K$_n$, and another one with $(n+1)^2$ peaks, K$_{n+1}$, 
at roughly $H/L\simeq (H_n+H_{n+1})/2L=1.18n+0.41$, i.e. at 
$H/L=1.59$, $2.77$, $3.95$, $5.13$, $6.31$, $7.49$, $8.67$, etc. 

This is shown in Fig. \ref{evol1}, where a sequence of structures (globally stable for each $H$) from K$_6$ 
to K$_7$ are plotted in local-density contour plots. In this case the transition is located somewhere between 
panels (c) and (d). At each transition point the two structures that coexist
will be slightly distorted: the one with $n^2$ peaks will be slightly expanded, whereas that with $(n+1)^2$ peaks
will be slightly contracted. The peaks of these structures are in fact a bit smeared out about the mean 
positions. The elastic free-energy cost associated with having a lattice parameter different
from that in bulk is the driving mechanism of the commensuration transitions. An animation showing different
commensuration phase transitions are included as a supplementary material to this paper.

\begin{figure}[h]
\includegraphics[width=5in,angle=0]{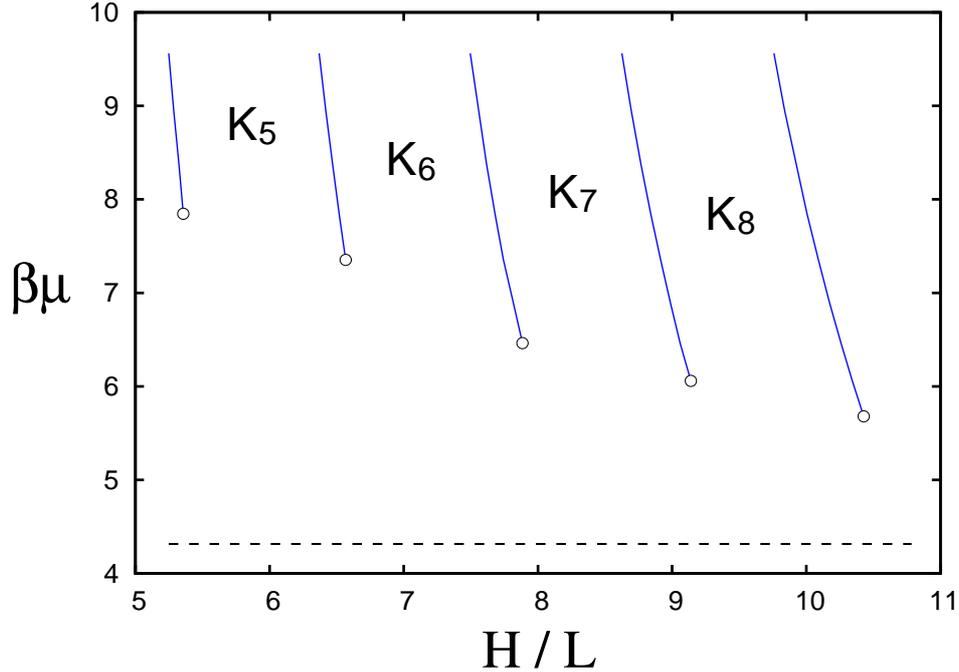}
\caption{Phase diagram of the HR fluid with aspect ratio $\kappa=1$ (hard squares) in the
chemical potential $\mu$ vs. cavity lateral size $H$. The open circles
indicate the terminal critical points. The dashed horizontal line is the bulk
chemical potential of the I$\to$C transition.}
\label{pd1}
\end{figure}

In Fig. \ref{pd1} the surface phase diagram is plotted in the $\mu$-$H$ plane.
The diagram covers a sequence of transitions from K$_4$ to K$_9$. The transitions terminate in critical points,
indicated by open circles in the diagram. The sequence of critical points, which are always above the bulk transition
(dashed horizontal line in the figure), tends to the bulk {I-(C or K) bifurcation value 
as the cavity size is increased. From a free-energy minimization using a Gaussian parametrisation of the density profile,
we obtain a bulk C-K transition located at a chemical potential $\beta\mu=10.096$, which is above the range shown in Fig. 
\ref{pd1}.} 
Also note that, since
the bulk transition is continuous, the transition in the cavity is suppressed because the system is confined in
both spatial directions and there can be no singularity in the free energy.

\begin{figure}[h]
\includegraphics[width=5in,angle=0]{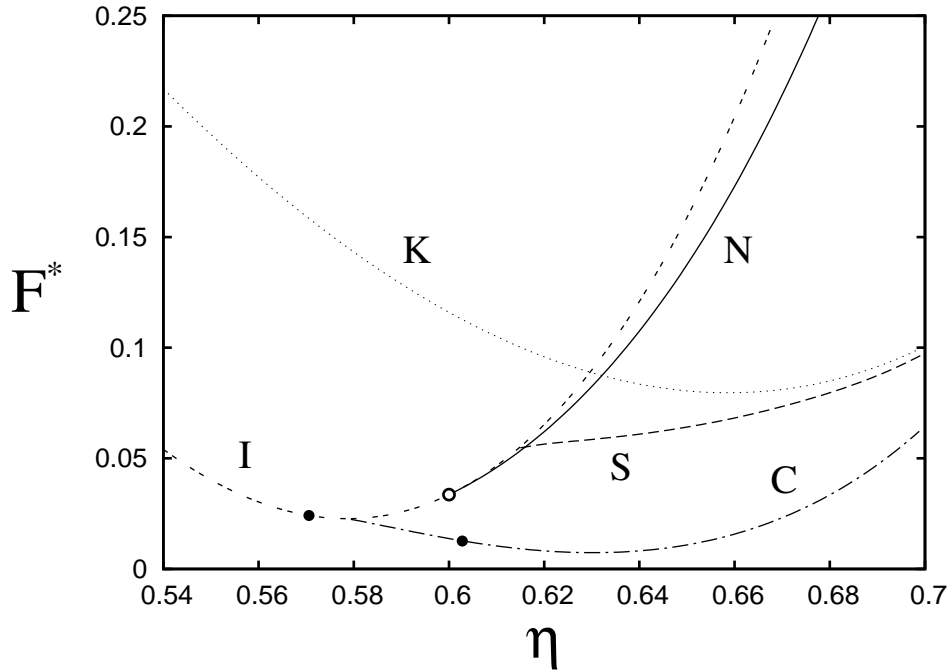}
\caption
{{Free-energy density branches in reduced units $F^*=\beta {\cal F}L\sigma/A$} 
of the I, N, S, C and K phases with respect to packing fraction for the HR fluid with aspect 
ratio $\kappa=3$. The straight line $f=a\eta+b$, with $a=7.17$ and $b=-3.91$, has been 
subtracted from the free energies in order to better visualise the curves. The coexistence packing fractions
for the I$\to$C transition are $\eta=0.571$ and $0.603$, and are represented by filled circles.
The bifurcation point for the I$\to$N transition (open circle) is at $\eta=0.600$.} 
\label{k3}
\end{figure}

\subsection{$\kappa=3$}

Now the sequence of bulk phases is
\begin{eqnarray*}
\hbox{I}\to\hbox{C}.
\end{eqnarray*}
The transition is of first order. The other phases, N, S and K are all metastable at least up to
$\eta\sim 0.73$ (the maximum density explored). Fig. \ref{k3} shows the free energies
of the different phases, which reflect the first-order character of the $\hbox{I}\to\hbox{C}$
transition; the packing fractions of the two coexisting phases are indicated by filled circles. 
We have not found a stable K phase with $\gamma_x\ne\gamma_y$ and $0<\gamma_x,\gamma_y<1$ (uniaxial K phase); 
thus the free-energy of the K phase plotted in Fig. \ref{k3} corresponds to that of a one-component fluid of parallel
HR (with $\gamma_x=1,\gamma_y=0$ or $\gamma_x=0, \gamma_y=1$). This solid 
is equivalent, after rescaling in the direction of the long rectangular axis, to a system of hard squares.
In Ref. \cite{Yuri} a plastic K phase with  $\gamma_x=\gamma_y$ was found as a metastable phase using the same model;
this branch is not represented here.

\begin{figure}[h]
\includegraphics[width=5in,angle=0]{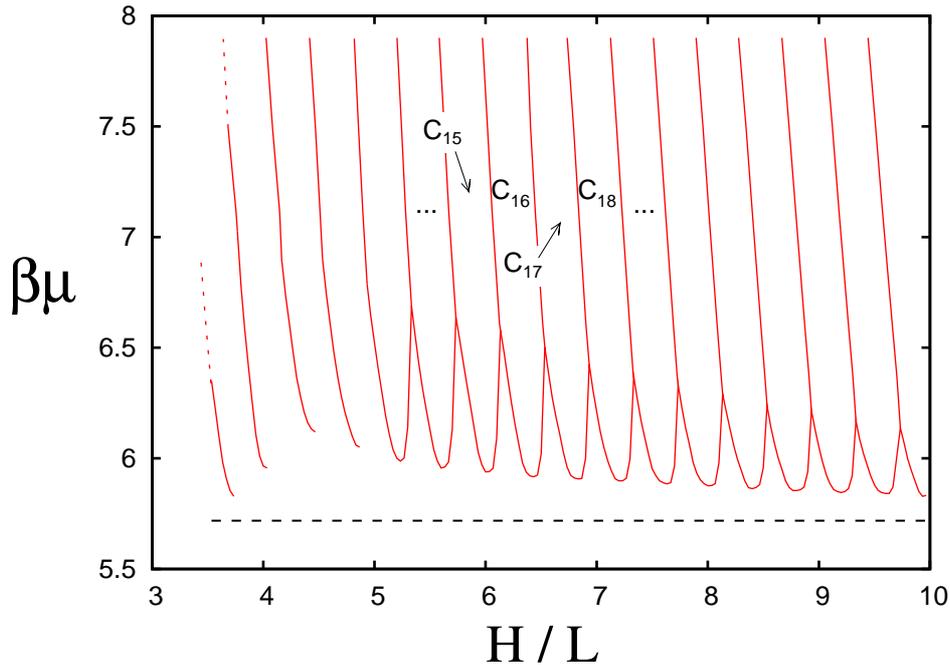}
\caption{Phase diagram of the HR fluid with aspect ratio $\kappa=3$ in the
chemical potential vs. cavity lateral size $H$. The dashed horizontal line is the 
chemical potential of the first-order bulk I$\to$C transition.}
\label{pd3}
\end{figure}

\begin{figure}[h]
\includegraphics[width=6in,angle=0]{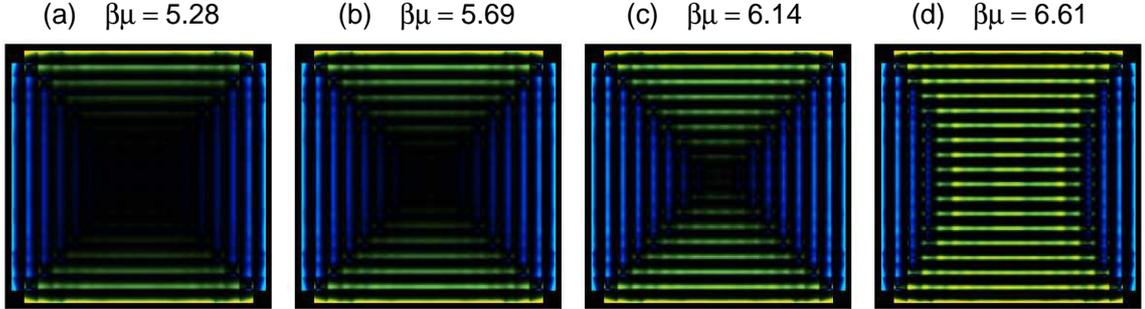}
\caption{Density false-colour plots of confined structures in the case $\kappa=3$ for fixed cavity size $H=7L$. 
Cases for different chemical potential values (in the range $\beta\mu=5.28$--$6.61$) are shown. Note that it is
actually the product $Q({\bm r})\rho({\bm r})$ [see Eqns. (\ref{pt}) and (\ref{Qr})] that is plotted in the
graphs, with the case $\rho_y>\rho_x$ represented in blue and the case $\rho_y<\rho_x$ represented in yellow.}
\label{evol3a}
\end{figure}

The surface properties of this fluid against a hard wall have been investigated in Ref. \cite{Yuri}
with the same theoretical model. The preferred orientation of the particles at the wall is 
parallel. This favours C-like configurations near the wall and, in fact, the C phase wets the wall-I interface, i.e.
as the transition density is approached from below, a film of C phase is adsorbed with a thickness
that diverges at the bulk transition. When the 
 system is confined in the cavity, the bulk $\hbox{I}\to\hbox{C}$ transition continues as a first-order
transition at chemical potentials {\it above} that of the bulk transition, see Fig. \ref{pd3}. 
In a wetting situation one would expect the transition to occur {\it below} the bulk transition. Here we are dealing with 
a frustration effect induced by the four surfaces. The transition line becomes
a highly nonmonotonic function of cavity size $H$ and, in fact, is connected to the
commensuration transitions that take place between different C-like structures. 
In Fig. \ref{evol3a} density false-colour plots are shown for different structures along a path at fixed $H=7L$
that crosses one of the I$\to$C transition curves. The structure is at first symmetric but with considerable
C-like oscillations propagating from the four walls. At the transition the symmetry is broken
and a well-developed columnar structure (horizontal columns in the central region), parallel to two of the surfaces,
appears in the cavity, with small islands of columns, in the perpendicular direction, adsorbed on the other two
surfaces. The confined C phase has a complicated structure that results from the inability
to satisfy the surface parallel-orientation at the four walls of the cavity with a single uniform C phase. 
Only two such conditions can be verified, and the result is the formation of two regions at two opposing walls
where surface orientation is parallel but opposite to that of the central columnar-like region.

\begin{figure}[h]
\includegraphics[width=6in,angle=0]{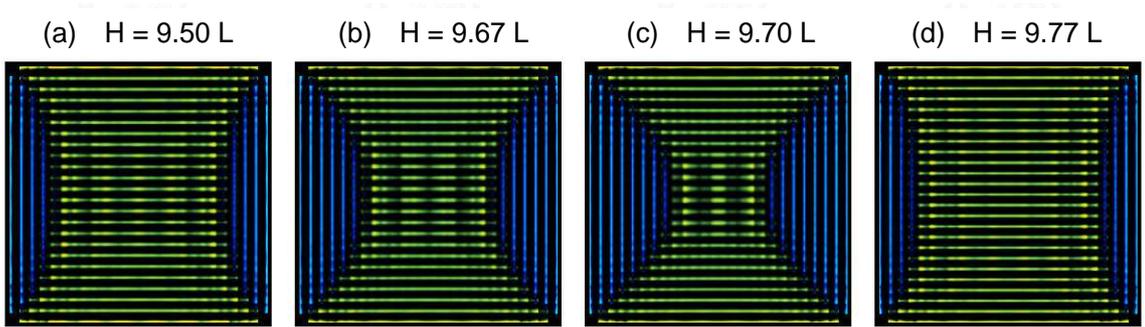}
\caption{Density contour plots of confined structures in the case $\kappa=3$ for $\beta\mu=7.901$, indicating 
structural changes across one commensuration transition. Different cavity sizes from
$H=9.50L$ to $9.77L$ are shown. These structures correspond to the C$_{24}\to$C$_{25}$ transition.
See caption of Fig. \ref{evol3a} for an explanation of colour code.}
\label{evol3b}
\end{figure}

The commensuration transitions are of the type C$_n\to$C$_{n+1}$, where C$_n$ is a columnar phase with $n$ columns. 
At the transition the system develops an additional column in the central region. Because the orientational order
is very high, the transition mainly involves translational degrees of freedom in the direction {\it perpendicular} to the
columns, and it is the wall distance along this direction that is relevant. Fig. \ref{evol3b} shows a sequence of 
configurations at high chemical potential, where different confined C structures are shown in the neighbourhood of the 
C$_{24}\to$C$_{25}$ transition. Panels (a-c) correspond to the free-energy branch of the C$_{24}$ phase. 
The thermodynamic transition occurs at $H=9.44L$ so that these panels correspond to metastable states. As $H$ increases, 
the two small regions with columns at perpendicular orientations grow in size at the expense of the central structure,
which shrinks and develops highly structured density peaks. Overall the structure gets more symmetric. 
When the system switches to the C$_{25}$ free-energy branch [configuration in Fig. \ref{evol3b}(d)], 
the peaks in the central region rearrange into a new columnar layer (with the surface structure largely unaffected) 
and the size of the regions with perpendicular orientation returns to its usual value.
Note that, in the stability window of the C$_n$ phases, these regions look very similar in structure and size 
regardless of the value of $n$ (for sufficiently large $n$). 
The sequence of configurations shown in Fig. \ref{evol3b} is not necessarily related to the 
actual kinetic behaviour of the phase transition, but gives an indication of how the nucleation of the new phase in
the cavity could take place. A film of the evolution of the fluid
as the cavity size $H$ is varied is presented as supplementary material to this paper.
Finally, the connection between the commensuration 
C$_n\to$C$_{n+1}$ and capillary $\hbox{I}\to\hbox{C}$ transitions is broken for small cavities,
which means that the I phase passes continuously into a C-like phase.

\subsection{$\kappa=6$}

\begin{figure}[h]
\includegraphics[width=5in,angle=0]{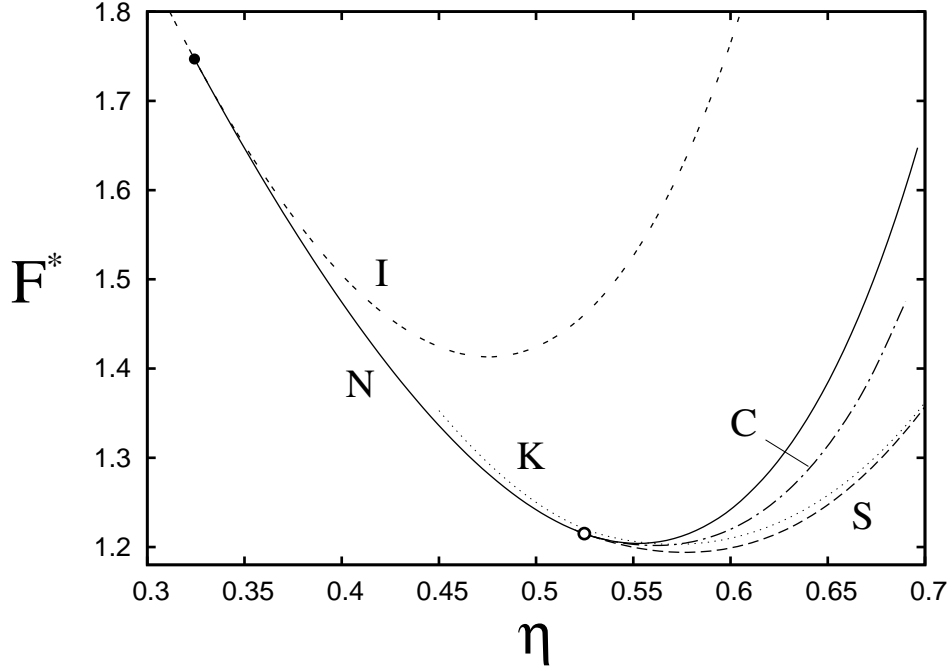}
\caption
{{Free-energy density branches in reduced units $F^*=\beta {\cal F}L\sigma/A$} 
of the I, N, S, C and K phases with respect to packing fraction for the HR fluid with aspect 
ratio $\kappa=6$. The straight line $f=a\eta+b$, with $a=5.5$ and $b=-4.0$, has been 
subtracted from the free energies in order to better visualise the curves. The I$\to$N transition (filled circle)
occurs at $\eta=0.324$, while the N$\to$S transition (open circle) is located at $\eta=0.525$.} 
\label{k6}
\end{figure}

For this aspect ratio the bulk sequence is very different. Now the C phase is no longer stable,
and instead the S phase becomes stable. The phase sequence is
\begin{eqnarray*}
\hbox{I}\to\hbox{N}\to\hbox{S}.
\end{eqnarray*}
Fig. \ref{k6} presents the free energies branches of the different
phases. The $\hbox{I}\to\hbox{N}$ transition occurs at $\eta=0.324$ and is continuous.
The $\hbox{N}\to\hbox{S}$ transition is also continuous and takes place at $\eta=0.525$.
The K phase is unstable at least up to $\eta=0.7$ but in becomes more stable than the S phase at higher
densities {(again the free-energy density of the K phase plotted in Fig. \ref{k6} corresponds
to a system of parallel HR, i.e. to a one-component system; since we expect the fraction of the perpendicular
species to be negligible in a full calculation, both free energies should be almost identical).} 
Bifurcation points for the (continuous) I$\to$N and N$\to$S transitions are indicated by circles.
The S phase is stable at high densities, but the difference in free energy with the K phase decreases,
so that it is likely that a transition to the crystal takes place at higher densities.
 
Since the bulk phase diagram of this fluid involves three instead of two phases, the surface 
phase diagram in the cavity is more complex. Two regions of distinct fluid behaviour can be identified,
and they are covered separately in the following.

\subsubsection{I$\to$N transition under confinement}
\label{IN}

\begin{figure}[h]
\includegraphics[width=6in,angle=0]{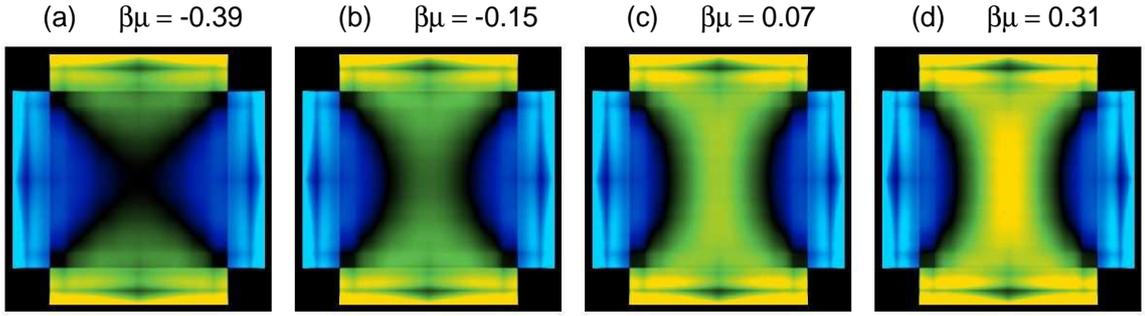}
\caption{Order-parameter false-colour plots of confined structures in the case $\kappa=6$ and for $H=3L$. 
Structures for different
values of chemical potential (indicated on top of each panel) are shown. (a) Symmetric configuration 
(confined I phase). (b-d) Symmetry-breaking configurations of increasing density.
The confined I$\to$N transition takes place between the (a) and (b) configurations.
The case $\rho_y>\rho_x$ is represented in blue, while the case $\rho_y<\rho_x$ is represented in yellow.}
\label{seq_IN}
\end{figure}

\begin{figure}[h]
\includegraphics[width=5in,angle=0]{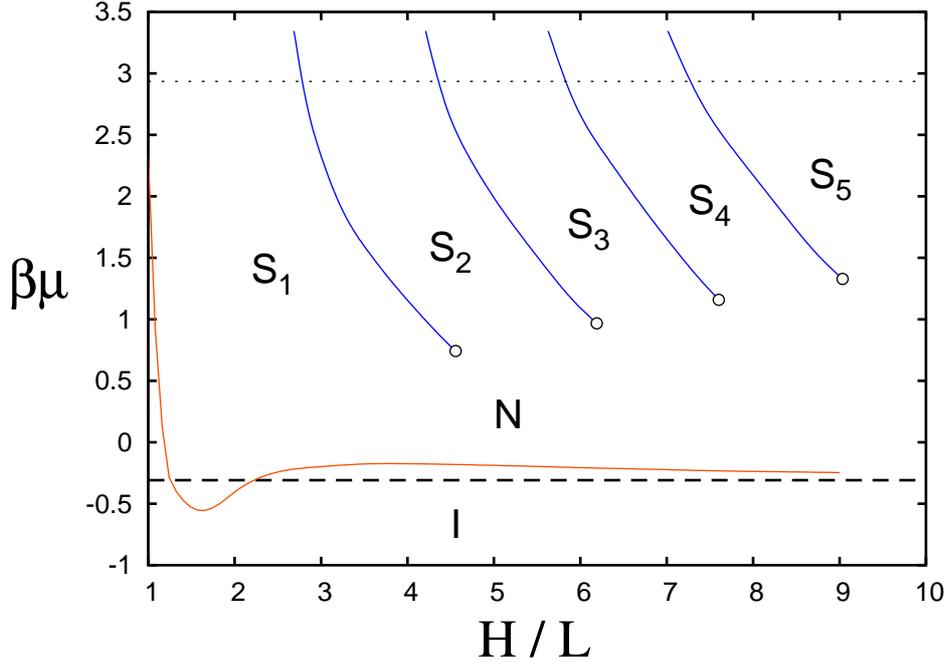}
\caption{Phase diagram of the HR fluid with aspect ratio $\kappa=6$ in the
chemical potential vs. cavity lateral size $H$.
Open circles are terminal critical points for the commensuration transitions. 
The dotted horizontal line is the chemical potential of the continuous N$\to$S transition, while
the dashed horizontal corresponds to the continuous bulk I$\to$N transition.}
\label{pd6}
\end{figure}

Since the bulk $\hbox{I}\to\hbox{N}$ is continuous, there should be no such transition under confinement.
The fact that we do find a transition line in the cavity that connects, for large cavities, with
the bulk transition, is due to the symmetry-breaking of the nematic director inside the square cavity. 
To see this, we study the
evolution of the structure inside the cavity as the density is increased at fixed $H$ from a low value.
We refer to Fig. \ref{seq_IN}, where the order-parameter field is plotted in false colour for the case 
$H=3L$. At low $\eta$ (or chemical potential $\mu$) the fluid is disordered (confined I phase), 
except at thin regions adsorbed at the four walls where
particles are oriented parallel (on average) to the corresponding wall, Fig. \ref{seq_IN}(a). 
This configuration is
fully symmetric since it conforms to the four-fold symmetry of the square cavity. At higher densities the 
fluid becomes globally oriented and the director breaks the symmetry by choosing one of two
possible but equivalent configurations (differing by a global rotation by 90$^{\circ}$). Figs. \ref{seq_IN}(b-d)
correspond to a choice where the director of the largest nematic region is oriented along the $x$ axis.
These two-fold symmetric configurations break the four-fold symmetry of the cavity. The
broken symmetry has an associated continuous transition, which occurs at a value of chemical potential
that tends to that of the bulk I$\to$N transition as the cavity becomes larger. The transition is shown in
Fig. \ref{pd6} (the phase diagram in the $\mu$-$H$ plane) as a continuous, nonmonotonic curve $\mu(H)$
in the neighbourhood of the dashed line (the bulk value of the chemical potential 
$\mu_{\hbox{\tiny bulk}}$). 

An interesting feature of the confined I$\to$N transition is that it may occur 
below or above the bulk transition depending on $H$. 
The behaviour of the confined I$\to$N transition for very large cavities can be understood from
the surface properties of the fluid near a single hard wall. Here we know that, as the 
continuous bulk I$\to$N transition is approached from below, there is critical wetting by the
N phase of the wall-I interface. In the cavity, close to but below the chemical potential of the
bulk transition, there should develop a film of the almost-critical N phase on each of the four
walls, causing orientational frustration in the central region of the cavity; this effect causes the 
symmetry-breaking mechanism to be postponed and the transition to occur above the bulk value
$\mu_{\hbox{\tiny IN}}$. 

For a cavity size $H=H_{\hbox{\tiny min}}\simeq 1.6L$ there is a minimum in the transition curve, followed by a 
maximum at larger cavities, $H_{\hbox{\tiny max}}\simeq 4.0L$. This feature is a consequence of particle 
size commensuration in the cavity. For $H\simeq 2.3L$ (where $\mu(H)\simeq\mu_{\hbox{\tiny IN}}$) 
the density maxima in the symmetric configuration occur near the four walls but no longer near the cavity
corners since not more than one particle can now accommodate parallel and 
close to the walls. When $H\simeq 1.2L$ (where again $\mu(H)\simeq\mu_{\hbox{\tiny IN}}$) 
the density maximum is displaced at the center of the cavity, particles find it more difficult to orient 
in the parallel configuration inside the cavity, and the transition line increases to very high 
chemical potential.



\begin{figure}[h]
\includegraphics[width=5in,angle=0]{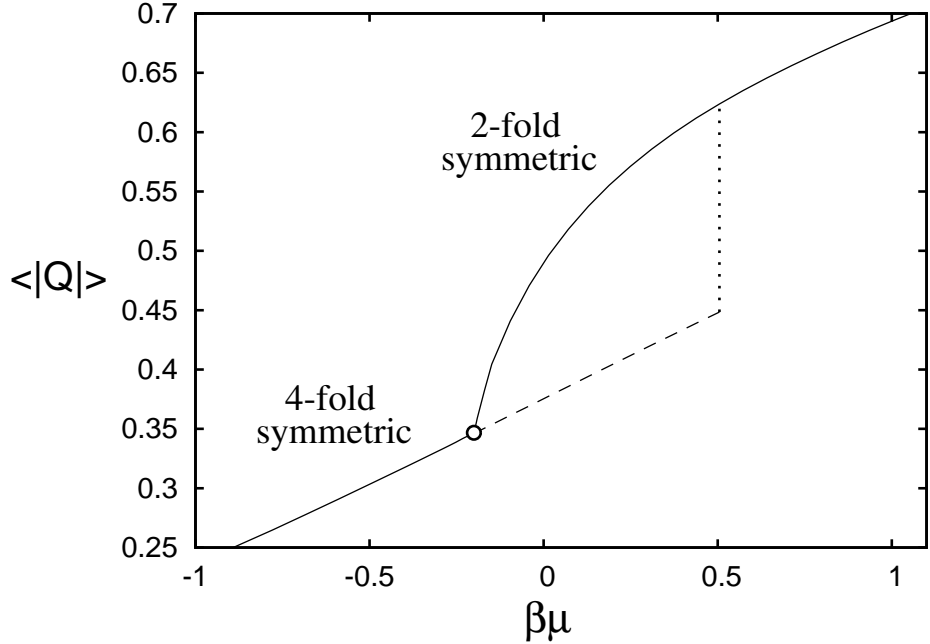}
\caption{Absolute value of the mean order parameter $\left<\left|Q\right|\right>$ 
of the HR fluid with aspect ratio $\kappa=6$ as a function of 
chemical potential $\mu$ and for $H=3L$. Continuous curve corresponds to the stable 
(lowest grand potential) solution. Open circle is the bifurcation point separating
the four-fold symmetric from the two-fold symmetry-breaking solution. Dashed curve is the continuation of
the (unstable) symmetric solution beyond the bifurcation point.
The dotted vertical line indicates the terminal point for the symmetric solution
and the jump to the symmetry-break since the smectic layers become more delocaliseding solution. Beyond the bifurcation point the
stable solution is doubly degenerate (two equivalent solutions rotated by $90^{\circ}$).}
\label{Q}
\end{figure}

The symmetric solution continues to be a solution (from
a numerical point of view) a bit beyond the bifurcation point. This is seen in Fig. \ref{Q},
where the integrated absolute value of the order parameter,
\begin{eqnarray}
\left<\left|Q\right|\right>=\frac{1}{A}\int_A d{\bm r}\left|Q({\bm r})\right|,
\end{eqnarray}
is plotted as a function of $\mu$ for 
$H=3L$. The four-fold symmetric solution (dashed curve) exists beyond the bifurcation point (open circle),
up to a point where it jumps to the two-fold symmetry-breaking, more stable solution.

Finally, the fact that the I$\to$N transition in the confined fluid is a symmetry-breaking transition
is confirmed by the fact that, when the cavity is slightly deformed into a rectangular shape,
the transition disappears altogether: all singularities in the grand potential vanish. 

\subsubsection{Smectic commensuration transitions under confinement}

\begin{figure}[h]
\includegraphics[width=6in,angle=0]{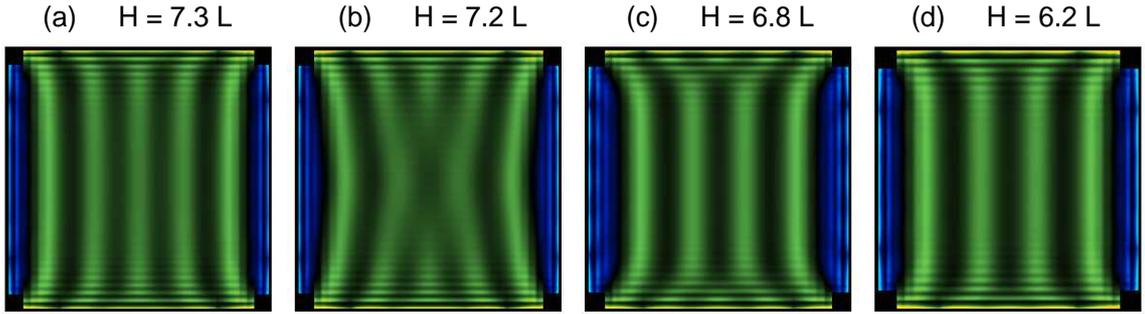}
\caption{Density contour plots of confined structures in the case $\kappa=6$ at high chemical potential,
$\beta\mu=2.336$. Different cavity sizes from
$H=7.3L$ down to $6.2L$ are shown. These structures correspond to the S$_{5}\to$S$_{4}$ transition.
Note that it is actually the product $Q({\bm r})\rho({\bm r})$ that is plotted in the
graphs, with the case $\rho_y>\rho_x$ represented in blue and the case $\rho_y<\rho_x$ represented in green.}
\label{evol6}
\end{figure}

As in the previous cases, the high density region of the phase diagram is dominated by commensuration 
transitions, but this time the transitions involve smectic phases with different numbers of layers. On increasing 
$H$ at fixed $\mu$, the system exhibits S$_n\to$S$_{n+1}$ first-order layering transitions 
between structures with numbers of layers differing by one (Fig. \ref{pd6}). These transitions are also obtained
at fixed $H$ for increasing $\mu$, and end in critical points 
at chemical potentials that lie below the bulk value $\mu_{\hbox{\tiny NS}}$ for the N$\to$S transition (dotted horizontal 
line in the figure), but that approach $\mu_{\hbox{\tiny NS}}$ as $H\to\infty$. This is a capillary effect: 
smectic layers are easily stabilised in the cavity since the order parameter rapidly saturates once the
symmetry-breaking nematic phase is established. Density plots for the transition between S$_4$ and S$_5$ are
shown in Fig. \ref{evol6} at fixed $\mu$ for decreasing $H$ (i.e. the system looses one smectic layer). 
Again, since the orientational order parameter is almost saturated, only the translational degrees 
of freedom are important; the commensuration mechanism is effective along the 
direction {\it perpendicular} to the layers and involves the distance between the walls parallel to the layers 
(vertical walls in the figure). The other two walls (perpendicular to the layers) induce increased ordering next to the 
walls. In contrast to the previous case, where the new column was nucleated in the central region and the surface
structure remained unaltered, in this case the 
layer-growth mechanism involves defects that begin or end at the walls perpendicular to the layers. In the case of layer growth,
two defects are created at the two walls which then propagate to the center of the cavity and give rise to a new smectic layer.
In the case where one layer disappears, the defects are formed at the centre and then migrate to the walls. 
This is the case in Fig. \ref{evol6}, which corresponds to the S$_5\to$S$_4$ transition as $H$ is decreased (see structures
for $H=7.2L$ and $6.8L$). Note that the structure next to the 
vertical walls grows a bit as the transition takes place. Again this behaviour may be representative of the nucleation processes
involving a change of one smectic layer. Panels (a) and (b) in Fig. \ref{evol6} correspond to the metastable S$_5$ phase;
in panels (c) and (d) the system is already in the stable free-energy branch of the S$_4$ phase.
A film of the evolution of the fluid as the cavity size $H$ is varied is also presented in this case as supplementary material.

\section{Limitations of the model and the I$\to$N transition}
\label{limitations} 

Our theory has a gross built-in approximation, namely,
particle orientations are restricted to only two perpendicular directions and this will certainly misrepresent
particle configurations where the sides of two particles are at an angle $0^{\circ}<\varphi<90^{\circ}$. 
By contrast, spatial correlations of parallel ($\varphi=0^{\circ}$) and perpendicular ($\varphi=90^{\circ}$) configurations 
are probably very well represented. Since the high-density bulk and confined phases are mostly composed of particles in
parallel or perpendicular configurations, 
the predictions of the theory on the bulk and surface properties 
in this r\'egime are qualitatively, if not quantitatively, correct.
This includes the prediction that, in bulk, the C or S phases are stable up to very high values of 
packing fraction, and that the C (S) phase is favoured for low (high) particle aspect ratios.
Phenomena involving the isotropic and nematic phases, although qualitatively correct, may be prone 
to larger errors. For example, the prediction of the 
suppression of the N phase in bulk for low aspect ratios is 
correct but the value of aspect ratio for which this occurs may not be very accurate, as well as
the location of the I$\to$N transition itself. An Onsager-like theory, such as the one used in 
\cite{delasHeras,delasHeras1}, would give more realistic predictions in this low-density region, but
will provide unreliable predictions in the high-density limit, which is the goal of our study.

\begin{figure}[h]
\includegraphics[width=5in,angle=0]{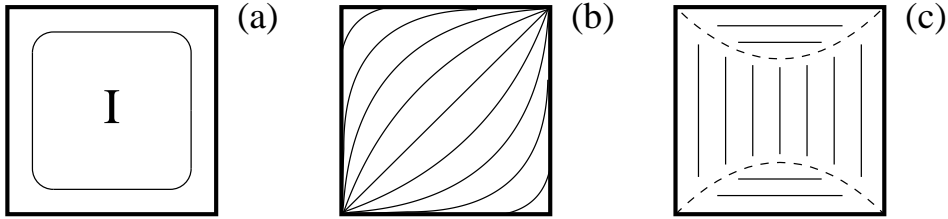}
\caption{Possible phases of the HR fluid confined in a square cavity in the nematic r\'egime. 
(a) Isotropic phase with
a film of nematic adsorbed on the inner walls. (b) Distorted nematic with two distinct pairs of 
defects, the components of each pair at opposite corners of the cavity. (c) Nematic with 
two opposing defect curves.}
\label{pha}
\end{figure}

Although our study is focused on the high-density region of the phase diagram, a side aspect is the nature
and structure of the confined nematic phase (in cases where this phase is
stable in bulk), and our model still could be useful well inside the nematic region. In our study we have always 
observed that the isotropic configuration, Fig. \ref{pha}(a),
changes to a nematic configuration where a region of uniform director is bounded by two smaller regions with 
the director at perpendicular directions (the structure discussed in
Section \ref{IN} and represented in the three right-most panels in Fig. \ref{seq_IN}). This configuration,
with its associated director field depicted schematically in Fig. \ref{pha}(c), 
satisfies the favoured orientation at the four walls with no elastic free-energy cost, but incurs 
a free energy due to the presence of two curved interfaces (domain walls) across which the director changes 
by 90$^{\circ}$. Experiments on quasi-monolayers of granular cylinders \cite{Galanis} show
the existence of a structure with two pairs of distinct defects located on opposite corners of the square 
cavity, Fig. \ref{pha}(b). In this structure the surface orientation is also satisfied, but there
are four point defects at the corners and some distortion of the director field, with the
corresponding elastic free energy. The stable equilibrium configuration will result, in a more realistic model,
from a balance between defect and elastic-distortion free energies. Since elastic constants increase with
density (see e.g. \cite{delasHeras1} for calculations on a two-dimensional model of hard discorectangles), the structure 
depicted in Fig. \ref{pha}(c) is expected to ultimately become stable at high packing fraction.  
Ongoing Monte Carlo simulations \cite{Dani} of the HR fluid confined in a square cavity show that, on increasing 
the density of the fluid, the isotropic phase becomes nematic following the phase sequence plotted in Fig. 
\ref{pha}. The fact that the intermediate structure depicted in Fig. \ref{pha}(b) is not observed in our model
is due to the simplistic representation of particle orientations which, as mentioned in Section \ref{FE},
precludes configurations where the director is distorted uniformly (i.e. without a region of discontinuity). 

{A way to systematically improve the present restricted model without dealing with the more numerically-demanding 
free-orientation model is to include more species in the set of restricted orientations. For example, 
by including species with orientations at $45^{\circ}$ and $135^{\circ}$ (and their equivalent orientations
at $225^{\circ}$ and $335^{\circ}$), particles could, for some external conditions, 
be highly oriented along the cavity diagonal and this constitutes a necessary (not suffient) condition to stabilize the 
structure shown in Fig. \ref{fig1}(e). Testing this hypothesis is difficult since the derivation of a fundamental-measure 
density functional for four species of HR would require severe approximations which 
would certainly affect the accurate description of particle correlations. As a first approach, we have checked that a 
Scaled-Particle Theory (SPT) description for the restricted-orientation model with four species gives a bulk I$\to$N 
transition curve located above the one obtained from the two-species SPT, which is the uniform limit of the present model 
(these results are not shown here). We mention that, as a bonus, the four-species SPT allows to describe the nematic phase with 
tetratic (four-fold) symmetry, a phase which is also predicted by the free-orientation models. This approach might be
worthwhile to follow in the context of future work on the structure of confined fluids.}

\section{Conclusions}
\label{conclusions}

In summary, we have studied the structure of a fluid of hard rectangles inside a square cavity,
using a fundamental-measure version of density-functional theory. 
Due to the restricted-orientation approximation inherent to the model, the prediction of the 
suppression of the N phase in bulk for low aspect ratios is 
correct but the value of aspect ratio for which this occurs may not be very accurate, as well as
the location of the I$\to$N symmetry-breaking transition itself. However, 
the prediction that, in bulk, the C or S phases are stable up to very high values of 
packing fraction, and that the C (S) phase is favoured for low (high) aspect ratios,
should be qualitatively, if not quantitatively, correct.

In the confined system the model predicts the occurrence of commensuration transitions 
between structures that differ in one unit cell (either C, S or K phases). 
The symmetry-breaking I$\to$N transition is obtained as a continuation of the bulk transition in
the confined system whenever there is a stable bulk N phase, 
and results from the square symmetry of the cavity; in rectangular, even slightly nonsquare, cavities, the I$\to$N transition
is suppressed altogether. The phase boundary of
the transition has a complicated, nonmonotonic behaviour with respect to cavity size; this behaviour is
probably correct for small cavities where, due to the square symmetry of the cavity, parallel and 
perpendicular orientations may be much more probable and the theory should be more accurate.

The general scenario that emerges from the present work is the following. In a severely restricted
geometry such as the square cavity, a liquid-crystal fluid is subject to several competing 
mechanisms, i.e. surface interaction causing frustration, elasticity and defect formation, the competition of which
causes a complex behaviour in the confined nematic. This problem has been studied several times
in the past. Since our model can predict the stability of
nonuniform bulk (S, C and K) phases, we have been able to extend these studies to incorporate
the effect of periodicity and the commensuration problems associated with a
periodic bulk phase in a confined geometry. Capillarity, surface-generated 
frustration and commensuration effects
all work together to create complex phase behaviour in the confined fluid.

Finally, several lines of future research may be worth pursuing. For example, a mixture inside the cavity 
adds a further mechanism in the way of demixing; the connection of bulk demixing and surface segregation with confinement, 
capillarity and frustration may add up to the richness in the phenomenology, with possible implications for
experiments on granular mixtures of particles. On the other hand, simulation studies of this system would be much welcome 
\cite{Dani} since they could hopefully confirm the behaviour implied by the present theoretical model. Also, the present model 
can be modified a bit to adapt it to a cylindrical geometry by introducing periodic boundary conditions in one of the
directions; this model could be useful to understand the bacterial growth mechanism recently suggested by Nelson and Amir
\cite{Nelson}. 

\acknowledgments
We acknowledge financial support from programme MODELICO-CM/S2009ESP-1691 (Comunidad Aut\'onoma de Madrid, Spain), 
and FIS2010-22047-C01 and FIS2010-22047-C04 (MINECO, Spain).

\end{document}